\documentclass[a4paper,12pt]{article}
\usepackage{amsmath,amssymb}
 \usepackage{epsfig} \usepackage{floatflt}
\usepackage{wrapfig}  \usepackage{subfigure}
 \input{DVabb11.sty}
 \graphicspath{{./pics-Slav70bis/}}

\def\Black{}

\begin{document}

 \begin{center}
\centerline\textsf{\large Dream-land with Classic Higgs field,
 Dimensional Reduction and all that\footnote{D.\,V. Shirkov,
 ``Dreamland with classic Higgs field, dimensional reduction and all that'',
 Proc. of the Steklov Inst. of Math., 2011, v.\,272, pp.216--222}}\bigskip

  D.\,V. Shirkov \medskip

 {\small Bogoliubov Lab., JINR, Dubna }
 \end{center}
  \centerline\textsf{Instead of Abstract} \medskip

  {\small This text, on the one hand, is related to the talk
  delivered at the Conference ``Gauge Fields. Yesterday, Today, Tomorrow'',
  dedicated to the Andrej Slavnov 70th anniversary;
  on the other --- in the form of a fairy tale --- it summarizes some results
  of researches performed after this Fest,
  mainly due to discussion around the talk.}
 \bigskip

  \section{\large Introduction}
 \subsection{\ns Motivation}
 Initial impetus for treating the topic mentioned
 in the title is related to the hot quest of the
 Higgs particle still escaping from observation.
 Anticipating the case with no Higgs peak detected
 at all, we turn to possibility that Higgs is a
 classical field (being an analog of Ginzburg--Landau
 order parameter in the theory of superconductivity)
 with nonzero constant component sufficient for the
 mass production in the current version of SM.

   However, changing a quantum Higgs for the classical
 external field yields the renormalization trouble in
 the EW sector of SM. Not entering this problem we looked
 for temporary practical remedy. The possible way is to
 use an invariant regularization procedure. To this goal,
 we exploit a transition from the four-dim space-time
 manifold to the one with a \ \textit{smaller} \ number of
 dimensions $\,D= 1+d<4\,$ at high enough energy.
 Technically, this provides us with one additional
 parameter, artificial cutoff scale or the
 range of reduction. The trick with changing the number
 of dimensions is a frequent one in literature (on
 superstrings etc.) devoted to the HE behavior. This
 Kaluza--Klein scenario with larger number of dimensions
 $\,D>4\,$ confronts us with the non-renormalizability.
 Instead, we consider another, an opposite possibility.\smallskip

 In Section 2, we use a $g\,\varphi^4\,$ QFT model with running
 coupling defined in both the two domains of $D=4\,$ and $D=2\,$;
 the $\gbar(Q^2)\,$ evolutions being duly conjugated at the
 reduction scale $\,Q\sim M=M_{dr}\,.$ Beyond this scale, in deep
 UV \ 2-dim region, $\gbar$ does not increase any more and tends
 to a finite value $\gbar_2(\infty)\,<\,\gbar_2(M^2)\,$ from above.
  As a result, the global evolution picture looks quite peculiar
 and can provide a basis for the modified GUT scenario with
 dimensional reduction for unification instead of leptoquarks
  ---  see Fig.\,8.\smallskip

 Then, Section 3 summarizes some very recent research that
 contains the study of Klein--Fock--Gordon eq. on toy examples
 with variable 2-dim space geometry including reduction of the
 space dimension number from $d=2\,$ to $d=1\,.$ Here, the
 specific trick of transforming the Klein--Fock--Gordon problem
 on variable geometry to Schr\"odinger-type eq. with potential
 generated by space variation was used.  \smallskip

  The final Section 4 contains a short summary and outlook.
 For a detailed exposition we address to fresh publications
 --- \cite{alisa} (on Section 2) and
 \cite{fiz+sh10,fiz+sh11} (on Section 3).

 \subsection{\ns Dimensional reduction}
 The diminishing of dimension number, \textit{dimensional
 reduction} (=DR) was used about 15 years ago (Aref'eva
 \cite{irina}, and others) in the HE scattering context.
 More recently, it got a second wind in the quantum gravity.

 To explore some practical aspects of the DR, we turn here
 to a limited subject, the transferring the renorm-invariant
 running coupling $\,\gbar(q^2)\,$ through the region of
 reduction and relating its behavior in two domains
 with different dimensionality. \smallskip

 In performing the DR procedure in the Lagrangian
 approach and in the energy-momentum integration of
 Feynman integral we use assumption \vspace{-3mm}

  \begin{quote}{\small\bf DR Agreement.} \ \textit{ Reduction at the space-time scale \
 $x_{dr}\sim 1/M_{dr}\,$ is, in a sense, equivalent
 to reduction at the energy-momentum scale} \
 $\,p_{dr}\sim M_{dr}\,.$ \end{quote}\vspace{-3mm}

 In the course of the first approach we use a sharp
 conjunction as an approximation to a softer mechanism
 of a continuous DR in the second one. \medskip

 {\ns\bf Classical illustration.} To illustrate the
 idea of approximation, take an empty wine bottle
 posed vertically like one presented on Fig.\,1(a).
\begin{figure}[h]
 \centerline{\includegraphics[width=0.65\textwidth]{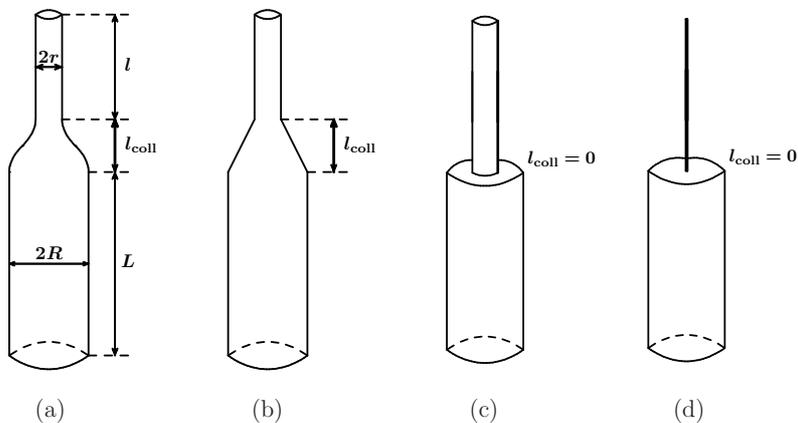}}
  \vspace*{-1mm}
   \caption{\small\label{fig1}
   (a) -- Wine-bottle with a long body and a long neck;\
   (b) -- Bottle with a conical collar: \
   (c) -- Bottle with a horizontal collar: \
   (d) -- Bottle with a horizontal collar and a thread-like neck.}
\end{figure}
 It has a main cylindrical body 
 with a relatively large radius $R$ and a
 length $L\,.$ The bottle's neck 
  of length $l\,$ and smaller radius $\,r$ is
 connected with the main part by a ``collar'' --- a
 transition region $\,C\,$ of varying radius and a
 short length $\,l_{coll}\,.$

 Imagine now an equation(s) describing some process on
 the 2-dim surface $S_2=S_{R,L}+S_{r,l}+S_{coll}\,$ of
 the bottle. One can mean a stationary boundary value
 problem, some dynamical process like wave propagation
 (with radiation) or heat conductivity and so on. A
 number of problems with exact analytical solutions on
 the surfaces $S_{R,L}\,$ and $S_{r,l}\,$ of cylindrical
 parts can be found. 
 Many of them could be solved for
 the whole two-dimensional manifold for simple enough
 forms of the junction collar region $C\,.$

  In the further analysis, along with smooth transition
 in Section 2.1, we shall use ``hard conjunction'' of
 two regions with different dimensions. The first one
 is an analog of Fig.\,1(a) in the limit $r\to 0\,,$
 while the second resembles the Fig.\,1(d).

 \section{\large Effective \gbar \ for the
  $\,\varphi^4\,$ model in various dimensions}    

  Take the one-component scalar massive quantum field
 $\varphi(x)\,$ with the self-interaction
  \beq\label{eq1}
 L=T-V;\quad V(m,g;\varphi)=\frac{m^2}{2}\,\varphi^2+
 \frac{4\pi^{d/2}}{9}\,g\,\varphi^4\,;\qquad g > 0\eeq
 in parallel in four ($D=4\,$) and two ($D=2\,$)
 dimensions.

 Limit ourselves to the one-loop approximation level
 for \gbar \ that corresponds to the only Feynman
 diagram contribution, the first correction to the
 4-vertex function, Fig.\,2.
 \begin{figure}[!ht]\centering
 \epsfig{scale=0.75,figure=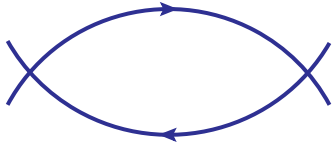} \vspace{-1mm}
 \caption{\small One-loop vertex diagram}
 \end{figure}

 Its subtracted contribution $\,I\,$ enters the effective
 running coupling as a whole :
  \begin{equation}\label{eq2} \gbar(q^2)=
 \frac{g_{i}}{1-g_{i}\,I\,(q^2; m^2, m_i^2)}\,.\eeq

 \subsection{\ns Smooth \ DR \ in the momentum picture}
  For DR in momentum representation 
 we modify the metric 
 $$ d k=d^4\,k\to d_Mk\,=\frac{d^4\,k}{1+k^2/M^2}\,;
 \quad\quad k^2=\mathbf{k}^2 - k_0^2\,\,. $$
 in the one-loop Feynman integrand Fig.\,2
 {\small  \[
 I\left(\frac{q^2}{m^2}\right)\,\to\,\frac{i}
 {\pi^2}\,\int\frac{d_Mk}{(m^2+k^2) [m^2+(k+q)^2]}
 =J(\kappa;\mu);\quad q^2=\mathbf{q}^2-q_0^2;\,\,
 \kappa=\frac{q^2}{4\,m^2};\,\,\mu=\frac{M^2}{m^2}.\]}
 Turn to UV asymptotics. In the ``deep 4-dim'' region
 $m^2\ll q^2\ll M^2$ one gets an ``intermediate''
 logarithmic behavior with the $M$ playing the role of
 the Pauli-Villars regulator. Meanwhile, in the ``deep
 2-dim'' region $q^2\gg M^2\gg m^2\,,$ the UV limit
 is finite. In usual normalization  \vspace{-4mm}

 \[ J \to J_i =  J(q^2/4m^2;\mu) - J(m_i^2/4\,m^2;
 \mu)\,;\quad m_i \sim m\,, \]
 one has \vspace{-6mm}

 \begin{equation}\label{uv-2} J^{[4]}_i(\kappa;\mu)\,
 \sim \ln\left(\tfrac{q^2}{m_i^2}\right)\,;\quad
 J^{[2]}_i(\kappa;\mu)\,\sim  \ln\left(\tfrac{4\,M^2}
 {m_i^2}\right)+\tfrac{M^2}{q^2}\ln\tfrac{q^2}{M^2}
 \,.\eeq

  The first expression is rising; while the second,
 decreasing. The maximum value of $J$ is attained
 at the DR scale $q^2 \sim M^2\,$ and is close to
 $\ln\mu\,.$ Hence, due to the DR, the evolution
 of the coupling $\gbar(q^2)\,$ changes; it
 diminishes beyond the reduction scale and tends
 to a finite value as in Fig.\,3.
  \begin{figure}[!ht] \centering
 \epsfig{scale=1.3,figure=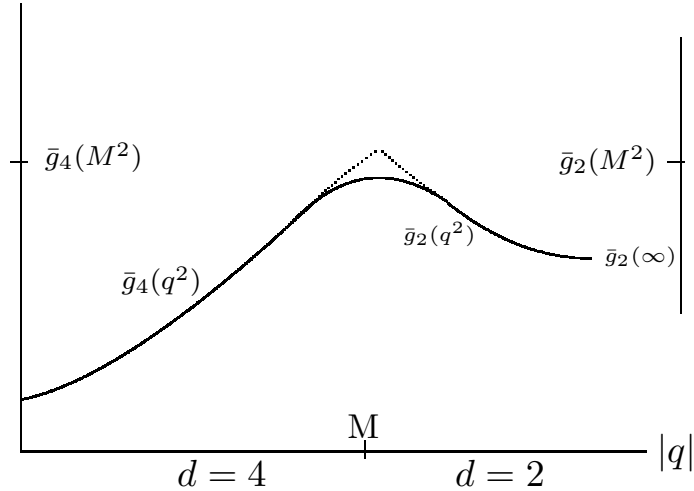} \vspace{-2mm}
 \caption{\small The  effective coupling evolution
 for the $\varphi^4\,$ model with DR.}\end{figure}

 \subsection{\ns Reduction with Lagrangian}
 Now, along with the DR Agreement, turn to the
 Lagrangian description eq.(\ref{eq1}). There,
 under transition to the $D=2\,$ case, the scalar
 field loses its dimensionality
 $\varphi_4(x)\to\varphi_2(x)\sim M\,\varphi_4(x)
 \,,$ while the coupling constant acquires it:
 $g_4\to\,g_2\sim M^{-2} g_4\,$ with one parameter
 that can be put equal to the DR scale
 $\,M=M_{dr}\,.$\bigskip

 \textsf{\ns Invariant coupling in $d=2\,$.} \
  In two dimensions, one can use finite Dyson
 transformations and formulate mass-dependent
 renorm-group (as it was first introduced\cite{nc56}
 in the mid-50s)
 \[\gbar^{[2]}(q^2)=\frac{g}{1-g\,I_2(q^2/m^2)}\,;
 \quad I_2\left(\frac{q^2}{m^2}\right)=\frac{i}{\pi}
 \int\frac{d^2k}{(m^2+k^2)\,[m^2+(k+q)^2]}\,.\]

 Here, $\,I_2\,$ is a finite contribution
 from 4-vertex diagram, Fig.\,2. It is a positive
 monoto\-nously decreasing function. Asymptotically,
 $I_2\sim\ln(q^2/m^2)/q^2\,,$ as in the
 second eq.(\ref{uv-2}). Hence, two-dim coupling in
 the UV limit tends to its limiting fixed value from
 above.\bigskip

 {\ns\sf Hard conjunction at the reduction scale.} \
 To obtain the joint picture of coupling evolution,
 consider transition from the ``low-energy'' 4-dim
 region \ $q^2<M^2\,$ to ``high-energy'' 2-dim one
 \ $q^2>M^2\,.$ For the ``hard'' conjunction
 (analog of Fig.\,1(d)) the continuity
 condition $\gbar_4\,(M^2)=\gbar_2(M^2)= g_M\,$
 yields the finite UV limit
 \beq\label{infty2}            
 \gbar_2(\infty)=\frac{g_M}{1+g_M\, I_2(M^2/m^2)}
 \,\,  <\,g_M\,.\eeq
 This means that above the reduction scale the
 effective coupling evolves from $g_M$ down to
 $\gbar_2(\infty)$ that corresponds to Fig.\,3.\Black

  \section{\large Klein--Fock--Gordon equation on variable geometry}
  Our attitude does not imply any modification of
 the concept of time. Instead, we have in mind some
 continuous transformation of spatial geometry up
 to reduction of space dimensions. \\
 To get physical intuition and experience, we start
 with Klein--Fock--Gordon scalar waves\cite{fiz+sh10}
 on some toy models of space (like on Fig.\,1) with
 variable geometry.

 Klein--Fock--Gordon equation (KFGEq) on a composite
 space can be presented in the form
$$ \Box\varphi-M^2\varphi=0\,;\quad
   \Box= - {\frac1{\sqrt{|g|}}}\,
            \partial_\mu
             \left(\sqrt{|g|}g^{\mu \nu}\,\partial_\nu\right)
       = -\partial^2_{tt}
       + \Delta_d\,,
$$ 
\ns with $d$-dim operator $\Delta_d$ on the $d$-dim metric $\gamma_{m n}$
$$ \Delta_d=\frac{1}{\sqrt{|\gamma|}}\,\partial_m
   \left(\sqrt{|\gamma|}\gamma^{m n}\,\partial_n
   \right)\,,
\quad m,n = 1,\ldots,d\,.
$$ 
 At the $(1+d)$-formalism with common time any
 global solution $\varphi$ on the parts with
 different dimension \ $d_1,d_2,\dots$\  have a
 common frequency $\omega=\omega_1=\omega_2=\dots\,.$

 On the cylinder-symmetry 2d-surface with \ shape
  function $\rho(z)$
$$ (dl)^2=\gamma_{mn}dx^mdx^n
         =\rho^2(z)\,d\phi^2
         +(1+{\rho^\prime}^2)\,dz^2\,,
$$
the Laplacian is \ 
$\Delta_2={\frac 1{\rho^2}}(
 \partial^2_{\phi\phi}+{\tfrac\rho {\sqrt{1+
 {\rho^\prime}^2}}} \,\partial_z{\tfrac \rho
 {\sqrt{1+{\rho^\prime}^2}}}\,\partial_z)\,.$\
Separating variables  $\varphi(t,\phi,z)=$
$=T(t)\Phi(\phi)Z(z)$ \ one gets two simple
  equations $T^{\prime\prime}+\omega^2 T(t)=0\,;\quad
  \Phi^{''}+m^2\Phi(\phi)=0\,,$ and the  more
  involved Z-equation \ \ \vspace{-6mm}

 $$\displaystyle
   \phantom{AAAAAaaa} 
   \frac1{\rho\sqrt{1+\rho'^2}}\,
    \partial_z
     \left(\frac\rho{\sqrt{1+\rho'^2}}\,\partial_z Z\right)
   + \left(\omega^2-M^2-\frac{m^2}{\rho(z)^2}\right)\,Z(z)
   =0, \hspace{17mm} (Z)
 $$
 with centrifugal potential $m^2/\rho(z)^2$ due to motion in curved space.

 Our next result can be formulated as \ {\it``Dynamics
 \ instead \ of \ Geometry''}. Indeed, Eq.(Z) can be
 transformed to a proper Schr\"odinger-like eq. :
$$\psi''(u)+\left(E-V(u)\right)\psi(u)=0$$
by change of coordinate \  and shape function
$$ z\mapsto u:\quad u(z)
 = \int\sqrt{1+\rho'(z)^2}\,
    \frac{dz}{\rho(z)}\,;
\quad \rho(z)\to \varrho(u)
 = \rho(z(u))\,.
$$
\begin{figure}[t!]
 \centerline{\includegraphics[width=15.truecm, viewport= -100 1 600 300] {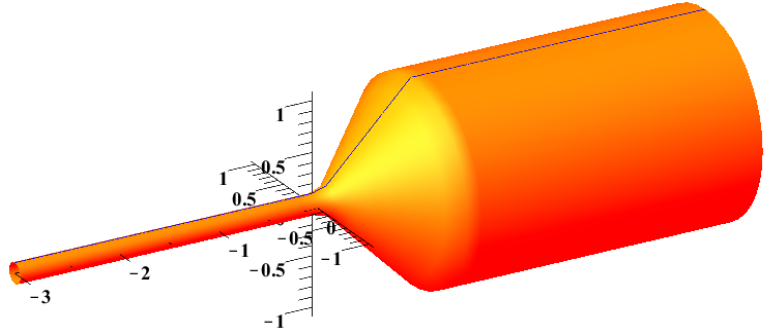}}
  \caption{\small The first toy model.\label{fig4}}
\end{figure}
\begin{figure}[!b]\vskip 2.truecm
 \centerline{\begin{minipage}{5.1truecm}\vskip .2truecm  \hskip -1.2truecm
  \includegraphics[width=2.4truecm,viewport=4 1 200 200]{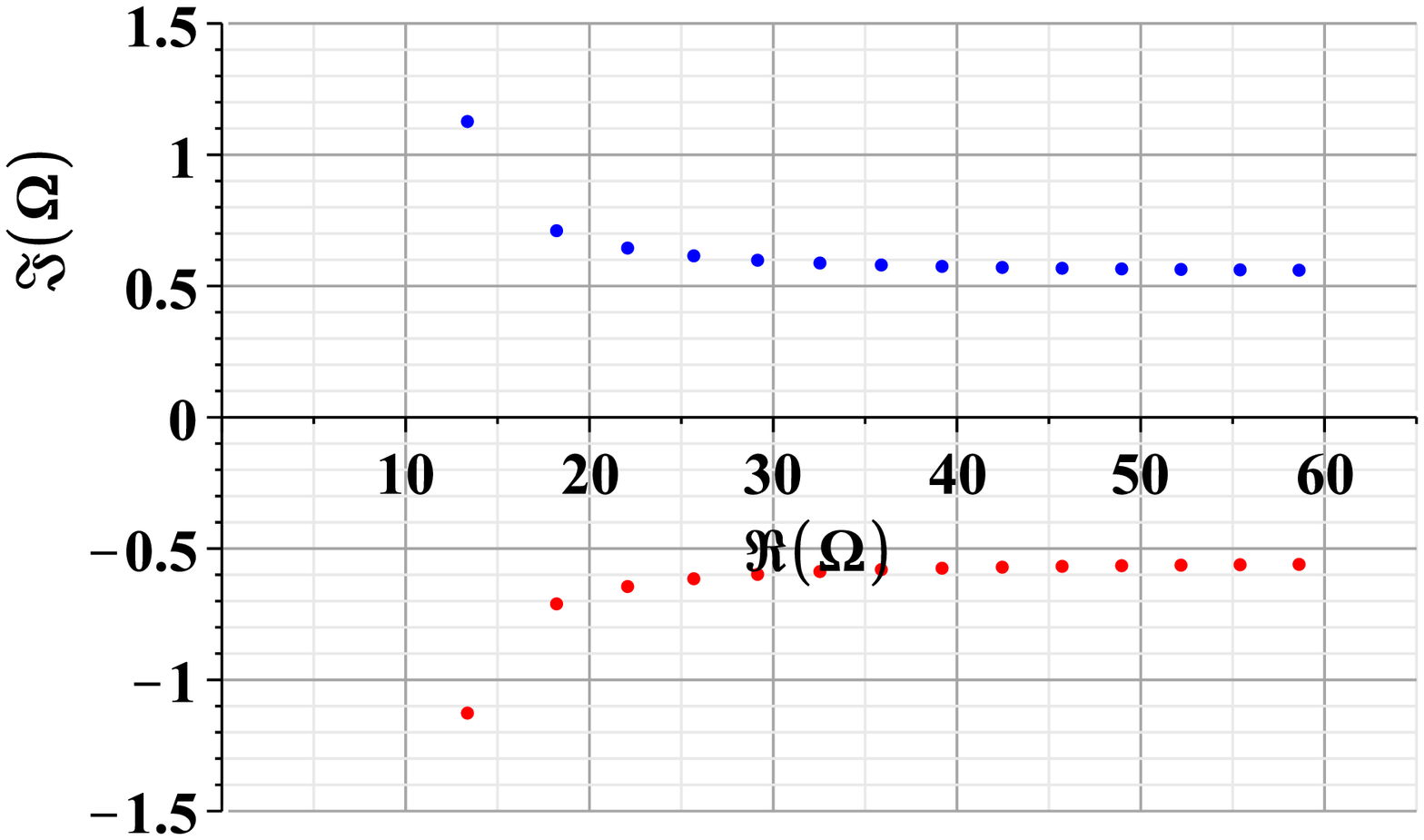}
  \end{minipage}\hspace{2.0truecm}\begin{minipage}{5.1truecm}\vskip -2.6truecm
  \hskip -1.truecm\includegraphics[width=2.3truecm, viewport=4 230 200 200]{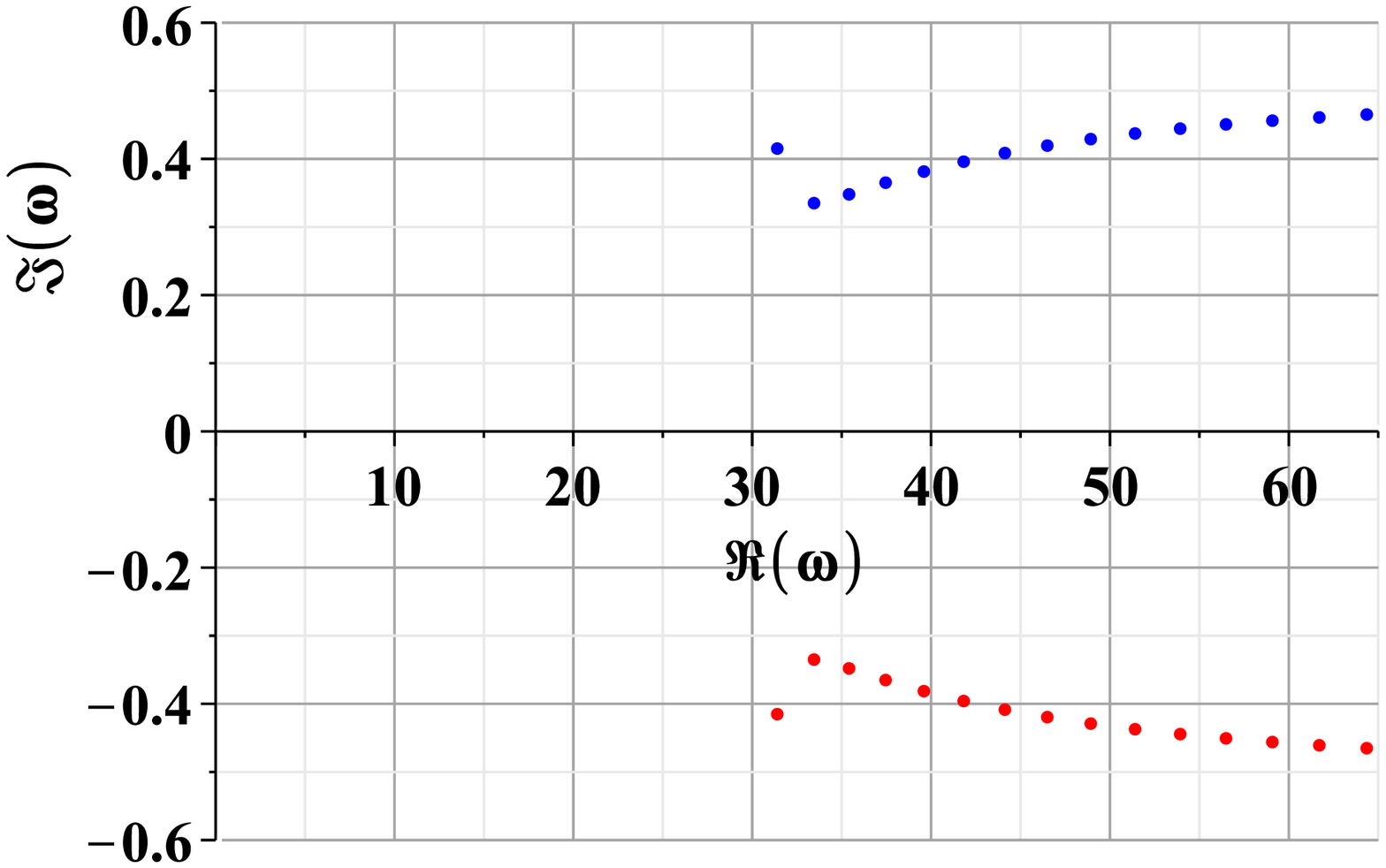}\end{minipage}}
  \caption{\small Complex spectra $\omega_{n,m}=
 Re \omega_{n,m} +i Im \omega_{n,m}, n=0,1,2,…$ for
 $m = 10, \alpha = \pi/3$ and masses $M=0$ (left panel)
 and $M=28.85$ (right panel) The sequence of points
 with Im$\omega_{n,m}>0$ corresponds to particles.
 The sequence of points with Im$\omega_{n,m}<0$
 corresponds to antiparticles.\label{fig5}}
\end{figure}
Then \
$\Delta_2=\varrho^{-2}\left( \partial^2_{\phi\phi}+
 \partial^2_{uu}\right)$ \
with $\varrho(u)^2$ \
being 2-dim conformal factor,
and
$$ E = 0,\quad V(u)=\left(M^2-\omega^2\right)
         \varrho(u)^2+m^2\,,
\quad\quad Z(z)=\psi(u(z))\,.
$$

Thus, study of KFGEq on a \ curved manifold \
is reducible to the Schr\"odinger-like eq. \
solving \
(with potential \ $V(u)$ \ defined \ by geometry).

Consider now KFGEq on surface of two cylinders
connected by part of cone with shape function
(Fig.\,\ref{fig4})
$$\rho_{cone}(z)=\left\{\begin{array}{cccc}
 R=\text{const}&:\hskip .2truecm \text{for}
 \quad z&\in& [z_R,+\infty), \cr
 z\cdot\tan\alpha&: \hskip .2truecm \text{for}\quad
  z &\in& [z_r,z_R],\cr
 r=\text{const}&:\hskip .2truecm \text{for}\,\,\,
  z&\in& (-\infty,z_r].
\end{array}\right.
$$

In the (singular) limit \ $r\to 0$ the model describes
dimension reduction $d=2 \to 1\,.$
It admits analytic solution in terms of Bessel functions
and getting spectra.
The related spectra depend on the KFG mass M,
as shown on Fig.\,\ref{fig5}.

Another analytically soluble case refers to conic-like
(with angle $\alpha\,$ at the top of the cone) junction
with thin cylinder and with smooth transition (ST)
to big cylinder
(Fig.\,\ref{fig6}).
It allows to find solutions
for class of Smooth Transition shapes \ $\rho_{ST}(z)$
and get spectra \ of KFG problem
including limiting DR case \ $r=0$.
\vspace*{-3mm}

\begin{figure}[th]
 \hspace*{33pt}\includegraphics[width=11truecm, viewport=-80 1 500 300]{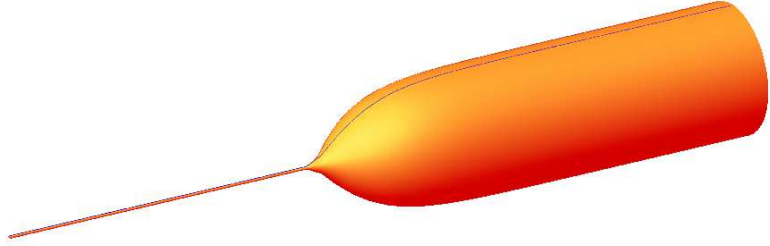}
  \vspace*{-1mm}
  \caption{\small Second toy model with angle $\alpha$ at the apex of the cone.
  \label{fig6}}
\end{figure}

  The spectra for ``2 Cylinders + smooth junction'' (Fig.\,7)
contain two series\footnote{
It is curious that 'minus solutions' \ $\omega^-_{n,m}$ \
are ``mounted from the cellar''.
In usual treating they are unphysical and omitted in text-books.}
of real frequencies $\omega_{n,m}$.
We show spectra for two values of the mass $M=1,\,\,2\,,$
the top angle value
$\sin\alpha=0.2\,$
and angular momentum \ $m=2\,.$
\vspace*{+5mm}

\begin{figure}[h]
 \hspace*{+13mm}
  \includegraphics[width=11truecm, viewport=-80 1 500 300]{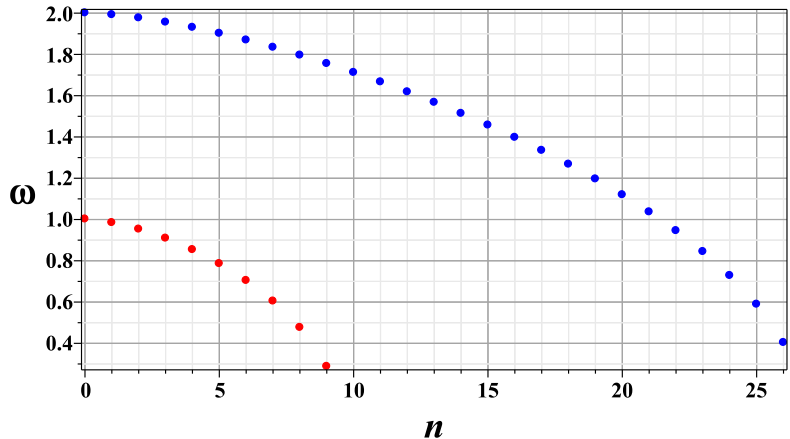}
   \vskip -.3truecm
  \caption{\small Real spectra for $\sin\alpha=0.2,\ m=2$.
   Ten red points correspond to $M=1$, whereas 27 blue points ---  to $M=2$.}
\end{figure}

 Comparing spectra of $\rho_{cone}$ case with various $\rho_{ST}$ ones
suggests a new idea:\\
One can try to generate and explain observed spectra of particles
using proper geometry of the junction
between domains of space
with different dimensions.

 \section{Conclusion}
   In the above analysis, one more alternative to
 the standard Higgs mechanism within the Standard
 Model was considered. The idea consists in
 employing the possibility of reducing the number
 of dimensions in the far UV limit.\bigskip

  \textbf{First,} taking \textit{ad hoc} the reduction
 from four (3+1) to two (1+1) dimensions at some
 high enough scale $|q|\sim M_{dr}\,,$ we studied
 the issue of effective coupling behavior for
 the $\varphi^4\,$ scalar model. 

\begin{figure}[!th] \centering
 \includegraphics[width=0.75\textwidth]{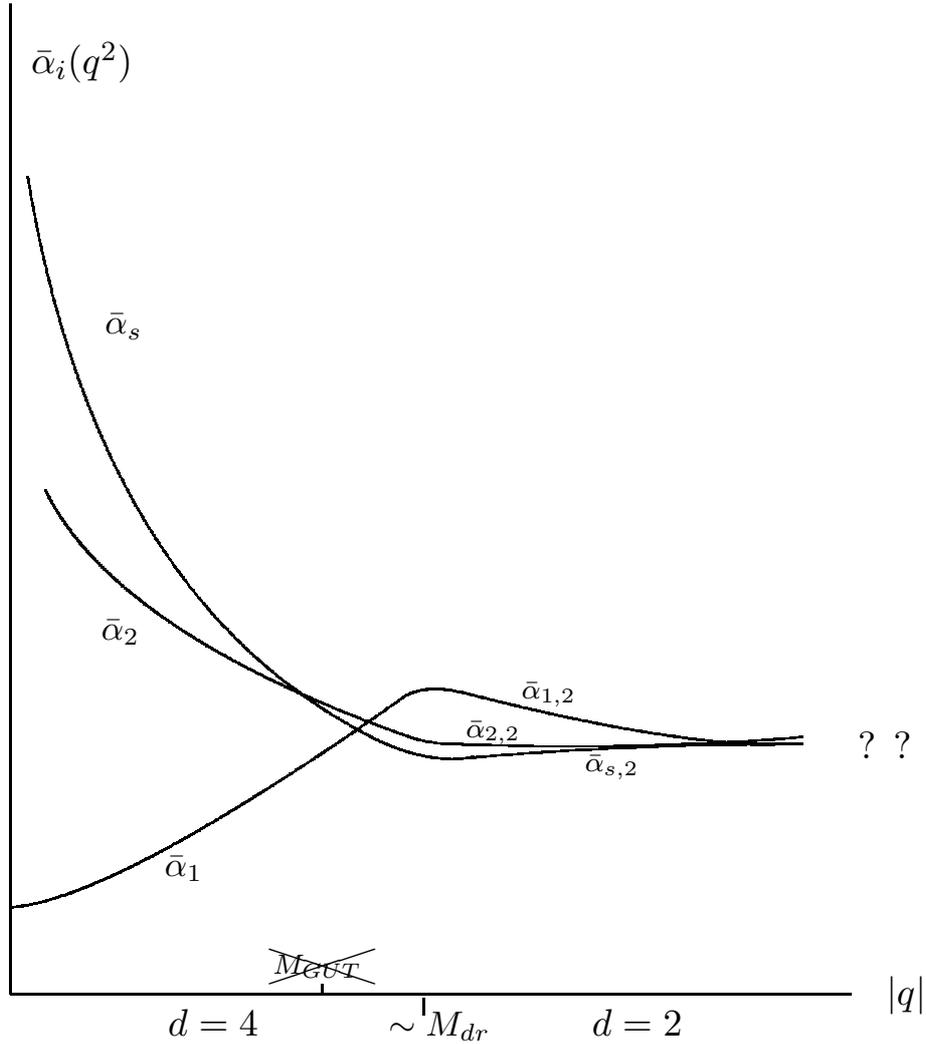}
  \caption{\small Modified chart \ of the \ GUT \ scenario
   provided by dimensional reduction instead of leptoquarks.}
\end{figure} 

   The peculiar property, the reverse running coupling evolution
 to a fixed point beyond the \ ``distorted looking-glass'' marked
 by the reduction scale has a chance to upgrade the Grand
 Unification scenario with the $M_{dr}\,$ value being of the
 order or even greater than the hypothetical
 lepto-quark scale --- see Fig.\,8.

  \textbf{Second}, we looked for the Klein--Fock--Gordon eq.
 solution on few toy models of space (like presented
 on Fig.\,1) with variable geometry. Here, the results
 are
\begin{enumerate}
 \item[\textsf{A.}] \textsf{On Signals between domains with \ Variable \ Geometry:}
  \begin{enumerate}
   \item[a1.]  Communication between diverse dimension domains
   by wave signals with $m\neq 0$ \
   is impossible. 
   \item[a2.] In the $2$-dim domain for real frequency \ $\omega$ \
   there is a total reflection on the cone of the wave
   coming from \ $z=+\infty\,,$
   (with change of the scattered wave phase). 
  \end{enumerate}
 \item[\textsf{B.}]  \textsf{Novel Phenomena:}

  The shape of effective mass and life-time excitation spectrum
  depends on the geometry of junction between different flat spaces. \\
  Thus, spectrum is a specific \ ``fingerprints'' \ of junction region
  which characterizes its \ Geometry. \
  Starting from spectrum one can re-construct the shape of junction. 
  In acoustics this problem was posed by Lord Rayleigh (1877),
  then advanced by Hermann Weil (1911) and later by others.
\end{enumerate}
\medskip
\medskip
\medskip

\centerline{\textsf{\large Acknowledgments}}
\medskip

It is a pleasure to thank Drs. Irina Aref'eva and Plamen Fiziev
for numerous useful discussions.
The courtesy of Dr. Jean Zinn-Justin
for informing on relevant literature is also appreciated.
This research has been partially supported
by the presidential grant Scientific School 3810.2010.2
and by RFFI grants 08-01-00686, 11-01-00182
\medskip 


\begin{thebibliography}{99}
 \itemsep -0.5mm
\bibitem{alisa}
 D.~Shirkov,
\textit{Phys. Part. Nucl. Lett.} v.\,7: 379--383 (2010);\
 arXiv:1004.1510\,[hep-th].
\bibitem{fiz+sh10}
 Fiziev P.~P. and Shirkov~D.~V.,
 \textit{Theor. Math. Phys.} v.\,167: 680--691 (2011);\
 arXiv:1009.5309v2\,[hep-th].
\bibitem{fiz+sh11}
 Fiziev~P.~P. and Shirkov~D.~V.,
 \textit{J. Phys. A} v.\,45: 055205/1--15 (2012),
  doi:10.1088/1751-8113/45/5/055205;\
  arXiv:1104.0903v2\,[gr-qc].
\bibitem{irina}
 I.~Ya.~Aref'eva,
 \textit{Phys. Lett. B}, v.\,325: 171--182 (1994);\
  arXiv:hep-th/9311115.
\bibitem{nc56}
 N.~N.~Bogoliubov and D.~V.~Shirkov,
 \textit{Nuovo Cim.}, v.\,3: 845--863 (1956).
\end{thebibliography}
\end{document}